\documentclass[runningheads]{llncs}
\usepackage[T1]{fontenc}
\usepackage{graphicx}
\usepackage{url}
\usepackage{tabularx}
\usepackage{multirow}
\usepackage{array}
\usepackage{makecell}
\newcolumntype{C}[1]{>{\centering\arraybackslash}m{#1}} % centered, vertically & horizontally
\usepackage{pifont} % add in preamble

\usepackage{ifpdf}
\usepackage{caption}
\usepackage{amssymb}% http://ctan.org/pkg/amssymb
\usepackage{pifont}% http://ctan.org/pkg/pifont
  \usepackage{cite}
  \usepackage{hyperref}
\usepackage{graphicx}
\usepackage{amsmath}
\usepackage{acronym}
\usepackage{array}
\usepackage{mdwmath}
\usepackage{mdwtab}
\usepackage{xcolor}
\usepackage{multicol}
\usepackage{tabularx}
\usepackage{longtable}
\usepackage{multirow}
\usepackage{subfig}
\usepackage{algorithm}
\usepackage{url}
\usepackage{etoolbox}
\usepackage{stackengine}

\usepackage{hhline}
\usepackage{booktabs}
\usepackage{todonotes}
\usepackage{booktabs}
\usepackage{bbm}
\usepackage{algorithm}
\usepackage{algpseudocode}

\usepackage{stackengine}
\usepackage{amssymb}

\usepackage{url}
\begin{document}

\title{Mitigating S-RAHA: An On-device Framework to Prevent Forwarding of Re-Captured Images}

\author{Keshav Sood*\orcidID{0000-0002-2127-1438} \and
Iynkaran Natgunanathan*\orcidID{0000-0003-4215-000X} \and
Purathani Praitheeshan*\orcidID{0000-0003-0569-9925} \and Praitheeshan Kirupananthan** \orcidID{0009-0008-9662-5289}}
\authorrunning{Sood et al.}
\institute{*School of IT, Deakin University, Geelong, VIC 3217,
Australia\\
%{**Blockstars Technology Pty Ltd. QLD 4214, Australia.}\\
{** Redshield Security Pty Ltd, Sydney, 2000}\\
\email{\{s224289198, keshav.sood, iynkaran.natgunanathan\}@deakin.edu.au}\\
{\email {purathani@gmail.com, praitheesh@gmail.com}}}
\maketitle

\begin{abstract}
Protecting sensitive visual content from unauthorized redistribution is a growing challenge for privacy-focused mobile applications, including dating platforms. %, and secure communication tools. %
%, law-enforcement sharing systems, and secure communication tools. 
Screenshot-prevention mechanisms, rely on server-side monitoring or are limited to digital screenshot detection, are commonly deployed to stop forwarding sensitive images. However, an adversary uses another smartphone to take a photo of the mobile-screen, in this scenario the existing solutions offer no protection against psychically screen-recapture attacks. Since the attack happens in the physical plane rather than on a digital plane and shows a void/hole in the existing solutions, we name this the \textbf{S}creen\textbf{-R}ecaptured 
\textbf{A}nalog \textbf{H}ole \textbf{A}ttack (\textbf{S-RAHA}). Such physically recaptured images bypass digital safeguards and can be freely forwarded, creating substantial privacy, personal-safety, and forensic risks. We present a low computational secure-by-design on-device framework that aims to detect and prevent the forwarding of recaptured images directly to the user’s device. The proposed system integrates a deep learning–assisted recapture detection model capable of distinguishing original digital content from camera-to-screen captures under diverse environmental conditions, together with an on-device enforcement mechanism that automatically blocks the sharing of suspected recaptured images between applications. We also introduce the concept of an invisible metadata identifier (IMI) that can be embedded into protected images to enable forensic traceability of potential leakage paths. Although the IMI component is explored at a conceptual and feasibility level rather than fully implemented, it demonstrates a promising direction for integrating lightweight, invisible identifiers into client-side security architectures. %Unlike existing approaches that rely on server-side monitoring or are limited to digital screenshot detection, the proposed framework provides a zero-trust, client-side solution that functionally restricts the spread of recaptured content. 
%Experimental results confirm that the deep learning module accurately identifies recaptured imagery while maintaining low computational overhead, highlighting the potential of the proposed architecture for enhancing visual privacy in modern mobile ecosystems.
%\end{abstract}

 \and screen-recaptured 
\end{abstract}

\section{Introduction}

Mobile devices have become the dominant medium for sharing visual content across social, professional, and personal communication platforms. Applications such as dating apps, telehealth systems, secure law-enforcement communication tools, and private social networks handle highly sensitive images. However, once displayed on a user’s screen, such images are vulnerable to redistribution. Even when an application disables screenshots or prevents direct forwarding, adversaries can bypass these protections through \emph{screen-recapture attacks}, where the displayed content is photographed using another mobile device. The resulting image is an ordinary photograph and can therefore be freely shared, creating significant privacy, personal-safety, and forensic risks. A high-level illustration of this scenario is shown in Fig. 1 below. We call it the S-RAHA. \par  

\begin{figure}[H]
  \centering
   \includegraphics[width=3.5in, height=2.1in]{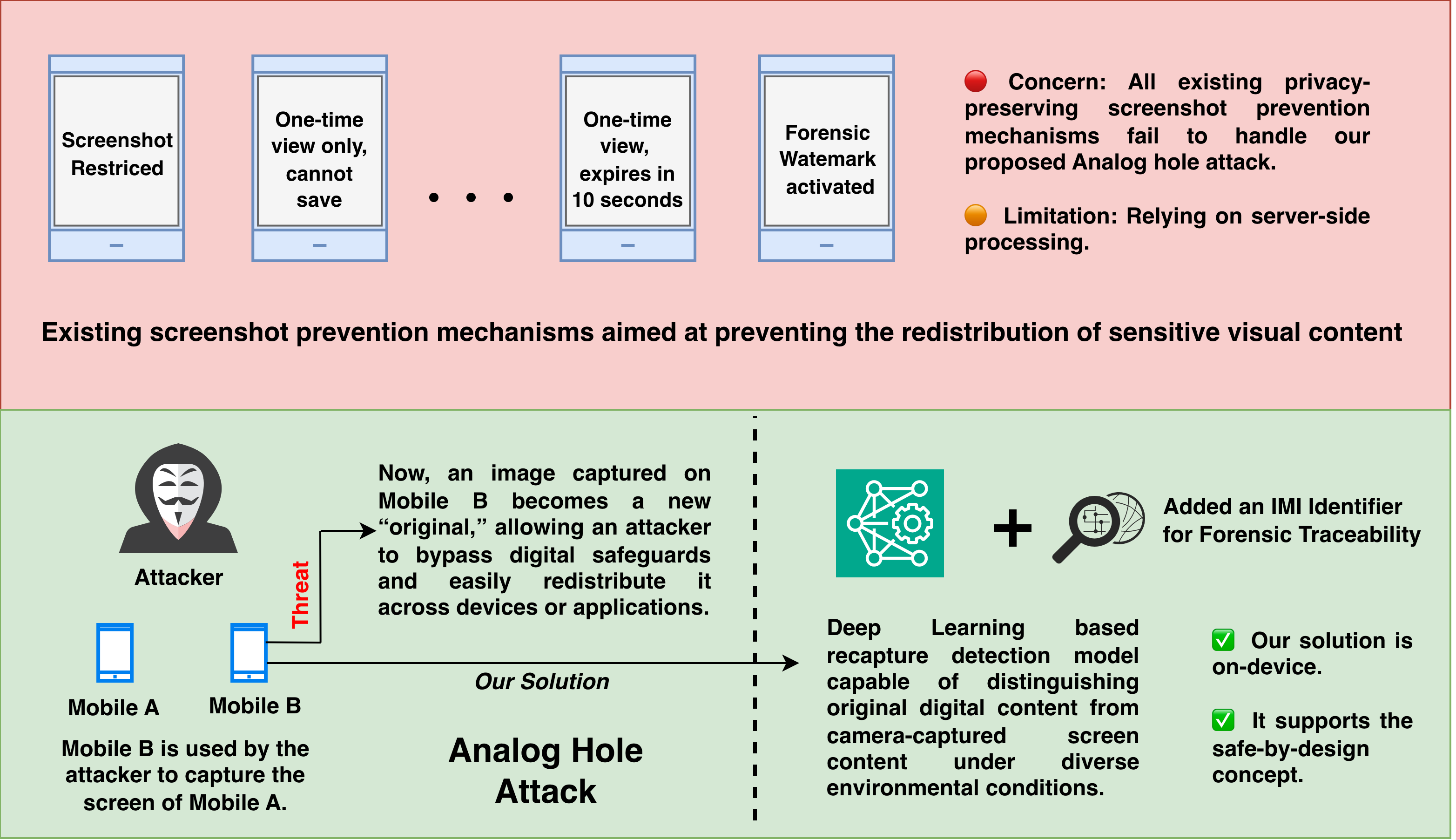}
  \caption{The proposed screen-recapture analog hole attack (S-RAHA).}
  \label{fig:framework}
\end{figure}
%%%%%
Early work in screen-recapture detection or image recapture-detection (IRD) has demonstrated that recaptured images exhibit distinctive artifacts caused by the display–camera pipeline, such as texture inconsistencies, illumination non-uniformity, and geometric distortions. Some commonly used approaches are based on moiré patterns~\cite{chen2025moire}, texture and blur analyses~\cite{ji2025towards}, chromaticity inconsistencies detections~\cite{chen2024cma}, specularity distribution checks, and illumination patterns check~\cite{li2025comprehensive}. Still, the current screen-recaptured image detection currently faces significant challenges as high-resolution 4K/OLED displays technology minimize the physical traces, like moiré patterns, that systems rely on for identification. These forensic tools often struggle with generalization. Also, the lack of diverse, high-quality training datasets and the advances generative tools used by attackers~\cite{park2025chimera},~\cite{279994} in which they intentionally add noise to mask recapture signatures, make it highly difficult to maintain high detection accuracy in general.\par

The work done by Ke~\textit{et al.} extracted multiple low-level statistical features including blur, color distribution, and texture variation to distinguish recaptured images from originals, achieving strong performance across different display and camera combinations~\cite{ke2013multiple}. Further, the authors~\cite{thongkamwitoon2015edge} advanced the field by learning discriminative dictionaries of edge profiles to improve robustness against variations in viewing conditions and device types. %Following these, from traditional signal processing techniques towards deep learning based approaches, 
More recently, domain-generalized and few-shot learning approaches have been proposed, including multi-scale residual feature extraction and attention-based fusion, enabling IRD models to generalize better to unseen devices and recapture settings~\cite{luo2021scale, hussain2025fewshot}. Dedicated smartphone recapture datasets have also been developed to support evaluation and training of modern IRD algorithms~\cite{dirik2011smartphone}. \par

Parallel research in digital watermarking aims to embed invisible, robust information into visual content for copyright protection, authentication, and forensic tracing. %Classical watermarking techniques operate in spatial and transform domains, while recent deep learning–based approaches significantly improve robustness against distortions. 
For instance, Bai~\textit{et al.} introduced SSDeN, a deep watermarking framework specifically designed to resist screen-shooting distortions by modeling realistic camera-to-screen noise~\cite{bai2022ssden}. Similarly, Cao \textit{et al.} proposed a universal deep watermarking method resilient to screen-shooting attacks and device-induced degradation~\cite{cao2024universal}. More advanced methods such as scale-adaptive U-Net watermarking with dynamic noise modeling have demonstrated strong robustness even under severe screen-cam distortions~\cite{liu2025screenshooting}. \par 

While these approaches~\cite{wang2023coarse},~\cite{chen2024cma},~\cite{ji2025towards},~\cite{park2025chimera},~\cite{chen2025unmask},~\cite{li2023recaptured} enhance post-incident detection and tracing, they do not directly prevent the redistribution of recaptured images (the attack shown in Fig. 1), leaving a gap for secure-by-design preventive mechanisms. %Further, existing protection mechanisms suffer from several limitations in the context of mobile applications. 
In addition, screenshot-prevention flags provided by mobile operating systems can be circumvented using an external camera; server-side monitoring is based on user compliance or network connectivity; and watermark-based forensic pipelines operate after the incident. %These limitations highlight the need for a~\emph{secure-by-design, client-side solution} that integrates physically recapture detection, enforcement, and forensic capabilities directly on the user’s device. 
%With this motivation, we propose a unified mobile architecture that combines recapture detection, on-device enforcement, and forensic traceability into a single client-side framework. 
Our contributions are as follows.
\begin{itemize}
\item 1) We revisit the analog hole attack problem in modern mobile ecosystems;  the problem is almost unexplored at all and we named it the  \textbf{S}creen\textbf{-R}ecaptured \textbf{A}nalog \textbf{H}ole \textbf{A}ttack (\textbf{S-RAHA}). No matter how many digital restrictions (screenshot blocks, one time view, etc.) are placed, an adversary can still successfully execute this attack. 
%For this scenario we have explored, software-based prevention (using invisible watermarking, hardware privacy filters, and presence detection) is extremely difficult since the attack happens in the physical world rather than the digital one. 
%We presented a deep learning--assisted recapture detection module capable of distinguishing digitally rendered images from camera-to-screen captures under varied lighting, angles, and device types. 
\item 2) We propose a secure-by-design, client-side (on-device) framework that integrates recapture detection, leakage prevention, and conceptual forensic tracing. The framework is robust and capable enough to distinguish digitally rendered images from camera-to-screen captures under varied lighting, angles, and device types. It provides an enforcement mechanism to automatically blocks forwarding of suspected recaptured images. 
%\item 3) We develop and evaluate a deep learning model for recapture detection tailored to mobile operating environments. 
A simplified version of the web application is presented to demonstrate its' technical feasibility. 
\item 3) We proposed a conceptual IMI and its feasibility as well for future integration which aims to support forensic traceability of leakage paths. %While the IMI component is explored at a feasibility level rather than a full implementation, it demonstrates how lightweight invisible identifiers can complement client-side controls.
\end{itemize}
\textit{Benefits:-} The proposed attack highlights the severity of the issue and motivates the need for effective mitigation strategies. The proposed secure-by-design solution brings its own advantages, i.e., a) mitigating the identified issue during the design phase and on-device is far more cost-effective, b) minimize potential entry points for attackers, c) easier to meet stringent standards such as GDPR, ISO 27001, and d) enhanced customer trust. %, e) the framework ensures that the added security layer protects the system leading to more stable and disruption-proof products. %The remainder of this paper is organized as follows. Section~\ref{sec:relatedworks} reviews related work from academic research as well as covers the industrial state-of-the-art. The section~\ref{sec:threat} presents the threat model and the problem formulation. Section~\ref{sec:methodology} explains the proposed framework and deep learning–based recapture detection model. Section~\ref{sec:web-application} provides the on-device enforcement mechanism. Section~\ref{sec:evaluation} presents experimental results. Section~\ref{sec:discussion} discusses limitations and future research directions. Finally, Section~\ref{sec:conclusion} concludes the paper.

\section{Related Work}
\label{sec:relatedworks}

We review academic work in interconnected areas: image recapture detection, digital watermarking for screen-shooting robustness, and mobile privacy controls.\par

\textbf{{Image recapture detection:-}}
Early work on recapture detection focused on handcrafted features exploiting artifacts from the screen-camera capture pipeline. Ke et al.~\cite{ke2013multiple} combined 136 low-level descriptors (blur, color distribution, texture) with an SVM classifier, achieving 97.2\% detection on a smartphone recapture dataset, though the low-resolution images used limit generalizability to modern capture scenarios. Further, Thongkamwitoon~\textit{et al.}~\cite{thongkamwitoon2015edge} argued that edge blurriness is the most scene-independent recapture cue and developed a dictionary-learning approach using K-SVD~\cite{aharon2006ksvd} to classify images based on line spread profiles extracted from sharp edges. Their method achieved $>$99\% true positive rate for recaptured images using only two features, representing a significant advance in parsimony. However, it relies on classical machine learning with hand-engineered features evaluated only on LCD displays under controlled conditions, and is not integrated into any enforcement pipeline.\par
More recently, domain-generalized and few-shot learning approaches have improved IRD robustness across unseen devices. Luo et al.~\cite{luo2021scale} proposed multi-scale residual feature extraction with attention-based fusion for better generalization to unseen display-camera combinations. Hussain et al.~\cite{hussain2025fewshot} framed IRD as a few-shot learning task, enabling reliable detection with scarce labeled data from new device pairs. Dirik et al.~\cite{dirik2011smartphone} released a dedicated smartphone recapture benchmark providing controlled references across handset and display combinations. While these methods advance detection accuracy, they remain standalone classifiers without client-side enforcement integration.\par

\textbf{{Deep watermarking robust approaches:-}} %A parallel line of research embeds invisible identifiers that survive the screen-camera distortion channel. 
Bai et al.~\cite{bai2022ssden} introduced SSDeN, a deep watermarking network trained with a differentiable screen-shooting noise model. Cao et al.~\cite{cao2024universal} generalized this with a device-agnostic training scheme, and Liu et al.~\cite{liu2025screenshooting} further strengthened robustness via a scale-adaptive U-Net with dynamic noise augmentation. \par 
\textbf{{Mobile privacy controls and their limitations:-}} Current mobile operating systems (OSes) offer screenshot-prevention flags (e.g., Android's \texttt{FLAG\_SECURE}) that block software-based capture but have no effect against physical recapture attacks. Server-side monitoring requires continuous connectivity and platform cooperation, while post-incident forensic pipelines are reactive. These limitations, together with the fact that existing approaches~\cite{wang2023coarse,chen2024cma,ji2025towards,park2025chimera,chen2025unmask,li2023recaptured} do not directly prevent redistribution of recaptured images, motivate the client-side, zero-trust architecture proposed in this paper. \par 
\textit{Key Gaps:-} These approaches are complementary but fundamentally different from our framework: they do not stop S-RAHA and they operate \emph{post-incident} for attribution, whereas our work also incorporates a conceptual watermarking-inspired component (IMI) as a forensic layer within a \emph{preventive} architecture.
In Table~\ref{tab:compare}, we discuss similar market product's overlapping capabilities and key differences; and the novelty of our work in Table~\ref{tab:compare2}.

\begin{table}[H]
\centering
\caption{A high level comparison with real market products}
\label{tab:compare}
\begin{tabular}{|p{0.14\linewidth}|p{0.30\linewidth}|p{0.30\linewidth}|p{0.30\linewidth}|}
\hline
%\centering \textbf{Product} & \centering \textbf{Description} & \centering \textbf {Overlapping capabilities} & \centering \textbf {Key differences} \\
%\hline

\multicolumn{1}{|c|}{\textbf{Product}} & \multicolumn{1}{c|}{\textbf{Description}} & \multicolumn{1}{c|}{\textbf{Overlapping capabilities}}  & \multicolumn{1}{c|}{\textbf{Key differences}}\\
\hline
VeraSnap v1.5 & Offering cryptographic evidence capture with multi-sensor fraud detection specifically targeting screen recapture. %The system employs barometric pressure attestation using Bosch BMP390 sensors, physiological tremor analysis (8-12Hz detection via 6-axis IMU), software-based screen recapture detection using tri-modal fusion (moiré pattern analysis, luminance distribution analysis, rolling shutter flicker detection) and NTP-based time consistency verification. 
& VeraSnap achieves 96\%+ detection accuracy with 5\% false positive rate, processing entirely on-device.  & %The fundamental difference lies in implementation approach, i.e., 
VeraSnap focuses on capture-time verification to prove content authenticity, while our proposal emphasizes post-capture enforcement to prevent unauthorized redistribution.\\
\hline
Truepic & It provides digital content authenticity infrastructure, the vision platform enables remote image verification. & Image authenticity verification, detection of manipulated content, and integration with mobile capture workflows. 
& Based on cryptographic signing and metadata analysis rather than recapture-specific detection algorithms, server-dependent verification model vs. client-side enforcement, proving authenticity rather than blocking redistribution. 
\\
\hline
DeepMedia AI & %It develops deepfake detection and media intelligence technology serving government agencies and enterprises. 
Detects digital harms including deepfakes, impersonation, and AI-driven misinformation across video, audio, and image media. %, with systems capable of processing approximately 10,000 videos per hour 
& AI-powered detection of manipulated imagery, real-time processing capabilities. 
 & Broader focus on deepfakes and synthetic media vs. specific recapture attack detection, primarily government contract-driven %(USD 1.25M Phase 2 SBIR from Department of Defence)
 rather than commercial mobile application integration, post-incident analysis orientation vs. preventive enforcement. 
\\
\hline
Magnet Forensics & Magnet Forensics provides digital investigation software for law enforcement, government agencies, and enterprises. %serving 4,000+ customers across 90+ countries. The company was acquired by Thoma Bravo for USD 1.3B in 2023. 
& Digital forensics and evidence analysis, image and video processing, law enforcement customer base. 
 & Desktop forensic workstation tools vs. mobile on-device solution, post-incident investigation focus rather than real-time prevention, comprehensive digital evidence platform vs. specialized recapture detection. 
\\
\hline
Amped Software  & Amped Software North America develops forensic software for image and video analysis in security and investigative sectors.%, serving law enforcement in over 80 countries.  
& Image enhancement and tampering detection, forensic analysis tools, law enforcement focus. 
 & Laboratory/workstation environment vs. mobile deployment, manual forensic examination workflow vs. automated enforcement, post-capture analysis vs. real-time blocking.  
\\
\hline

\end{tabular}
\end{table}

\begin{table}[H]
\centering
\caption{Novelty assessment of our solution}
\label{tab:compare2}
\begin{tabular}{|p{0.30\linewidth}|p{0.80\linewidth}|}
\hline
\multicolumn{1}{|c|}{\textbf{Criteria}} & \multicolumn{1}{c|}{\textbf{Description}} \\
\hline
Integrated detection and enforcement architecture & While multiple vendors offer recapture detection capabilities as discussed in Table~\ref{tab:compare}, %(VeraSnap, academic research achieving 97% accuracy using Vision Transformers), none combine detection with automated blocking mechanisms in a unified mobile framework. Existing solutions either:
they verify authenticity at capture time but do not prevent subsequent redistribution (exmaple VeraSnap),  they provide post-incident forensic analysis (Magnet Forensics, Amped Software), they rely on server-side monitoring requiring network connectivity (Truepic). Our proposal implements a zero-trust, client-side solution that both identifies recaptured content and blocks its' forwarding without requiring server validation or user compliance.\\
\hline
Mobile-first prevention vs. forensic analysis & The digital evidence management systems market focuses predominantly on post-incident investigation and evidence preservation. Companies like Cellebrite, OpenText, and CentralSquare provide platforms for collecting, storing, and analyzing digital evidence after security incidents occur. Our solution represents a paradigm shift toward preventive security by blocking unauthorized redistribution before it occurs, specifically addressing the threat model of privacy-focused mobile applications (dating platforms, secure messaging, law enforcement sharing systems).\\
\hline
Invisible Metadata Identifier (IMI) for traceability & The IMI concept introduces a novel forensic capability beyond binary detection. While the component remains at conceptual/feasibility stage, it addresses a critical gap: when recaptured content does escape, organizations need to trace leakage paths to identify compromised users or devices.
Existing watermarking approaches (forensic watermarking used by Irdeto for content protection) operate post-incident and require visible or detectable markers. The IMI concept aims for invisible, lightweight identifiers embedded directly into client-side security architectures.\\
\hline
Addressing the screen recapture attack vector & As noted in forensic research, recapture attacks defeat verification systems by creating genuine camera captures of manipulated content, erasing forensic traces that would normally reveal editing. The attack requires no sophisticated equipment — any smartphone can photograph a screen. Most authentication solutions focus on: screenshot prevention (bypass-able with external cameras), metadata consistency (eliminated through recapture), compression artifact analysis (reset by new capture process). Our solution specifically targets this overlooked attack vector with dedicated detection algorithms analyzing physical artifacts (moiré patterns, chromatic aberrations from display backlighting, focus distance characteristics, doubled tone mapping).\\
\hline
Commercial application integration gap  & While government and forensic markets are well-served by existing vendors (Magnet Forensics serves 4,000+ law enforcement customers; DeepMedia AI holds Department of Defence contracts), commercial mobile application integration remains under-served. Social media companies, dating platforms, and secure messaging services require lightweight, privacy-preserving solutions that do not introduce server dependencies or user friction. Our on-device architecture addresses regulatory compliance needs (the technology can help platforms be compliant with regulations) while maintaining user experience.\\
\hline
Remarks on the development stage and commercialization considerations & The proposed solution currently sits at concept validated in lab scale, with simulation testing completed. To further test it on real-world, ethics approval is needed for further analyses to validate the accuracy and efficiency of the proposal. The potential adopters including Australian Federal Police (and across the globe), government cyber forensic agencies, and social media companies seeking regulatory compliance. \\
\hline
\end{tabular}
\end{table}

\section{Threat Model and Problem Formulation}
\label{sec:threat}

\textbf{{Attack scenario:-}} Consider a mobile application geared toward privacy, such as a dating platform or a secure messaging system, where users exchange sensitive visual content under the assumption that it remains confined to the application environment. Modern mobile operating systems provide mechanisms (e.g., Android's \texttt{FLAG\_SECURE}) to prevent direct digital capture, but these are limited to \emph{software-based} capture and offer no defense against \emph{physical recapture attacks}, wherein an adversary photographs the displayed content using a secondary device. The proposed attack (S-RAHA) proceeds in three steps: (1)~the victim's device displays protected content ($I_{\text{orig}}$), (2)~the adversary uses a camera-equipped Device~B to photograph the screen of Device~A, producing $I_{\text{recap}}$, and (3)~$I_{\text{recap}}$, being an ordinary photograph with no forensic markers linking it to the source, is freely redistributed through any channel. This is particularly concerning because the recaptured image bypasses all application-layer controls, embedded watermarks or metadata may be degraded during the screen-camera process, and the adversary requires only commodity hardware.\par

\textbf{Adversary model:-} We consider an adversary with legitimate access to the device displaying protected content and a second smartphone for recapture. The adversary can control lighting, camera positioning, and viewing angles to optimize recapture quality, and may apply basic post-processing (brightness/contrast adjustment, cropping, rotation) before redistribution. The adversary is aware of screenshot-prevention mechanisms and seeks to circumvent them through physical recapture, but does not have white-box access to the detection model (black-box probing remains possible) and uses commodity smartphones rather than professional equipment. Critically, the screen-camera capture process inevitably introduces detectable physical artifacts---moir\'e patterns, illumination non-uniformity, geometric distortions, reduced sharpness, and color shifts---which form the basis for our detection approach.\par

\textbf{Security objectives:-} The framework aims to: (1)~accurately distinguish $I_{\text{orig}}$ from $I_{\text{recap}}$ under diverse conditions (lighting, angles, screen types, devices), (2)~automatically block forwarding of detected recaptured images client-side before they leave the device, (3)~enable forensic traceability via invisible identifiers surviving the recapture process, (4)~maintain usability with low latency and minimal false positives, and (5)~resist evasion attempts including post-processing, geometric transformations, and varied capture conditions.\par

\textbf{Problem formulation:-} Let $\mathcal{I}_{\text{orig}}$ 
and $\mathcal{I}_{\text{recap}}$ denote the distributions of 
original and recaptured images. Our goal is to learn a classifier 
$f: \mathbb{R}^{H \times W \times 3} \rightarrow \{0, 1\}$, 
where $H$ and $W$ denote the image height and width in pixels 
and $3$ represents the RGB colour channels, such that:

\begin{equation}
f(I) = \begin{cases}
0 & \text{if } I \sim \mathcal{I}_{\text{orig}} \text{ (original)} \\
1 & \text{if } I \sim \mathcal{I}_{\text{recap}} \text{ (recaptured)}
\end{cases}
\end{equation}
minimizing the expected risk, where $E$ denotes the average over the data distribution (expectation):
\begin{equation}
\mathcal{R}(f) = \mathbb{E}_{I \sim \mathcal{I}_{\text{orig}}}[\mathbbm{1}(f(I) = 1)] + \mathbb{E}_{I \sim \mathcal{I}_{\text{recap}}}[\mathbbm{1}(f(I) = 0)]
\end{equation}
subject to computational efficiency ($< 500$ms inference on mobile devices), generalization across unseen device combinations and conditions, and false positive control ($\text{FPR} < \tau_{\text{FPR}}$). Unlike traditional post-incident image forensics, our framework requires \emph{preventive detection} integrated directly into the on-device application workflow.

\section{The Proposed Framework}
\label{sec:methodology}
The proposed framework, see Fig.~\ref{fig:framework}, comprising: (i) a deep learning-assisted recapture detection module, (ii) an on-device enforcement mechanism, and (iii) a conceptual invisible metadata identifier for forensic tracing.% The framework operates through the following stages. 
\begin{figure}[H]
  \centering
   \includegraphics[width=4.5in, height=3.4in]{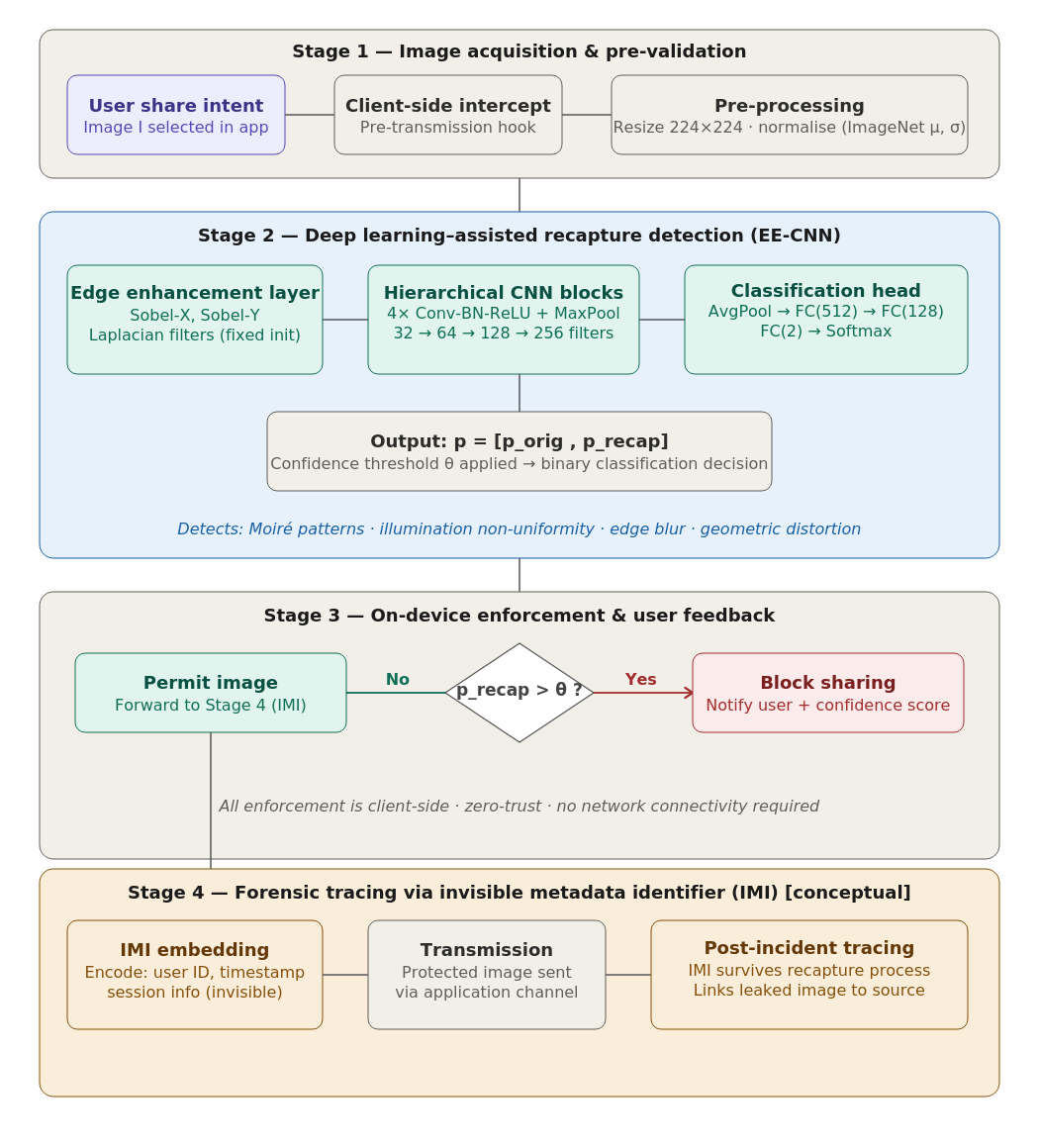}
  \caption{A high-level view of the proposed secure-by-design on-device framework.}
  \label{fig:framework}
\end{figure}
%The framework operates through the following stages:\par

\textbf{Stage 1: Image acquisition and pre-validation:-} When a user attempts to share an image $I$ within the protected application, it is intercepted by the client-side validation module using RESTFul API and preprocessed (resized to $224 \times 224$ pixels, normalized) for neural network inference.\par
\textbf{Stage 2: Deep learning-based recapture detection:-} The preprocessed image tensor is fed to an Edge-Enhanced Convolutional Neural Network (EE-CNN) trained to distinguish original from recaptured images. The model outputs a probability distribution $\mathbf{p} = [p_{\text{orig}}, p_{\text{recap}}]$. If $p_{\text{recap}} > \theta$ (a confidence threshold), the image is flagged as recaptured.\par
\textbf{Stage 3: Enforcement and user feedback:-} If original ($p_{\text{orig}} > \theta$), the image proceeds and an IMI can optionally be embedded for forensic tracing. If recaptured ($p_{\text{recap}} > \theta$), sharing is immediately blocked with a notification including the confidence score. All enforcement occurs client-side, ensuring zero-trust operation independent of network connectivity.\par
\textbf{Stage 4: Forensic tracing (conceptual):-} For permitted original images, the system can embed an invisible metadata identifier encoding user ID, timestamp, and session information. If leakage occurs through a recapture that evades detection, the IMI enables post-incident tracing. This component is explored conceptually as a direction for future integration.\par
Table~\ref{tab:threat-coverage} summarizes how the framework addresses various threat scenarios. Algorithm~\ref{alg:enforcement} formalizes the complete detection and enforcement pipeline described in the above stages.\par
%\textbf{Key Design Principles:} The framework adheres to: \textbf{Client-Side Enforcement}---all detection and blocking occur on-device before network transmission; \textbf{Zero-Trust Architecture}---validation does not rely on user compliance or network policies; \textbf{Transparency}---users receive clear feedback with confidence scores when images are blocked; \textbf{Performance-Aware Design}---the lightweight CNN achieves $<$500ms inference on commodity mobile hardware; and \textbf{Defense-in-Depth}---combining preventive, reactive, and forensic controls.\par
\textbf{Integration with mobile applications:-} The framework integrates via a RESTful API endpoint (\texttt{/api/validate-image}) returning JSON validation results, with WebSocket support for real-time feedback. While our proof-of-concept  (PoC) uses a web-based architecture (React frontend, Express.js backend, PyTorch inference), the approach generalizes to native mobile platforms through TensorFlow Lite or Core ML. If the detection model is unavailable, the system defaults to a secure-by-default posture, blocking all uploads. Our approach leverages domain-specific knowledge of recapture artifacts through an Edge-Enhanced CNN that explicitly prioritizes edge information---a key discriminative feature between original and recaptured images.\par

%\textbf{Threat Mitigation Coverage:} 
%Unlike existing approaches relying on server-side monitoring or post-incident forensics, our framework provides \emph{proactive, on-device protection} that restricts the spread of recaptured content before redistribution.

\begin{table}[H]
\centering
\caption{Threat mitigation coverage}
\label{tab:threat-coverage}
\begin{tabular}{|p{0.35\linewidth}|p{0.65\linewidth}|}
\hline
\textbf{Threat} & \textbf{Mitigation} \\
\hline
Screenshot redistribution & Prevented by existing platform 
screenshot-prevention mechanisms and complemented by 
API-level validation. \\

\hline
Physical screen recapture & Detected via EE-CNN based on moir\'e 
patterns, edge artifacts, illumination non-uniformity. \\
\hline
Post-processed recaptured images & Robust feature extraction 
captures artifacts even after brightness/contrast adjustment. \\
\hline
Leakage through external channels & API-enforced blocking 
prevents recaptured images from entering the messaging 
pipeline. \\
\hline
Insider threats & IMI embedding enables forensic tracing of 
original images. \\
\hline
Model evasion & Mitigated through diverse training data; 
adversarial robustness is discussed further in 
Section~\ref{sec:discussion}. \\

\hline
\end{tabular}
\end{table}

\begin{algorithm}[t]
\caption{On-Device Recapture Detection and Enforcement}
\label{alg:enforcement}
\begin{algorithmic}[1]
\Require Image $I$ selected for sharing, trained EE-CNN model $\mathcal{M}$, confidence threshold $\theta$
\Ensure Decision $d \in \{\texttt{PERMIT}, \texttt{BLOCK}\}$

\Statex \textbf{Stage 1: Pre-Validation}
\State Validate file type $\in$ \texttt{image/*} and size $\leq$ 10MB
\State $I_r \leftarrow \text{Resize}(I, 224 \times 224)$
\State $I_n \leftarrow \text{Normalize}(I_r, \mu_{\text{ImageNet}}, \sigma_{\text{ImageNet}})$

\Statex
\Statex \textbf{Stage 2: Edge-Enhanced Recapture Detection}
\State $E \leftarrow \text{EdgeEnhance}(I_n)$ \Comment{Sobel-X, Sobel-Y, Laplacian}
\State $F \leftarrow \text{HierarchicalCNN}(E)$ \Comment{4 conv blocks: $32 \to 256$}
\State $\mathbf{p} \leftarrow \text{Softmax}(\text{FC}(F))$ \Comment{$\mathbf{p} = [p_{\text{orig}}, p_{\text{recap}}]$}

\Statex
\Statex \textbf{Stage 3: Enforcement Decision}
\If{$\mathcal{M}$ is unavailable}
    \State $d \leftarrow \texttt{BLOCK}$ \Comment{Fail-closed policy}
\ElsIf{$p_{\text{recap}} > \theta$}
    \State $d \leftarrow \texttt{BLOCK}$
    \State Notify user with confidence score $p_{\text{recap}}$
    \State Delete image from temporary storage
\Else
    \State $d \leftarrow \texttt{PERMIT}$
\EndIf

\Statex
\Statex \textbf{Stage 4: Post-Decision Processing}
\If{$d = \texttt{PERMIT}$}
    \State Embed IMI$(I, \text{userID}, \text{timestamp}, \text{sessionID})$ \Comment{Conceptual}
    \State Transmit $I$ via application channel
\EndIf

\State \Return $d$
\end{algorithmic}
\end{algorithm}

%\subsection{Deep Learning--Assisted Recapture Detection Model}
%\label{sec:experiments}

\textbf{Data collection and pre-processing:-} We constructed a balanced dataset of 1,500 original smartphone photographs and 1,500 recaptured images obtained by displaying originals on various screens (LCD and OLED) and re-photographing with different smartphones under varied conditions (indoor/outdoor lighting, angles $15^\circ$--$45^\circ$, distances 20--50cm, diverse scene content). Images are resized to $224 \times 224$ pixels and normalized using ImageNet channel statistics. Training augmentation includes random horizontal flips, rotations ($\pm 5^\circ$), color jittering ($\pm 10\%$), and Gaussian blur ($\sigma \in [0.1, 2.0]$).\par

\textbf{Neural network architecture:-} The EE-CNN comprises three components. The \textit{Edge Enhancement Layer} initializes the first convolutional layer ($16$ filters of size $3 \times 3 \times 3$) with classical Sobel and Laplacian edge detection kernels, with remaining filters learned during training. This design explicitly targets the blurred edges, moir\'e patterns, and high-frequency artifacts characteristic of recaptured images. The \textit{Hierarchical Feature Extraction} module consists of four convolutional blocks, each with two convolutional layers, batch normalization, ReLU activation, and $2 \times 2$ max-pooling, progressively reducing spatial dimensions from $224 \times 224$ to $14 \times 14$ while increasing feature depth from 32 to 256 channels. The \textit{Classification Head} applies adaptive average pooling, followed by two fully connected layers ($512 \rightarrow 128$ neurons) with dropout regularization ($p = 0.5, 0.3$), producing a two-class softmax probability distribution over original and recaptured classes.\par

\textbf{Training configuration:-} The model is trained using cross-entropy loss with Adam optimizer ($\alpha = 10^{-4}$) and L2 weight decay ($\lambda = 10^{-4}$), running for 50 epochs with early stopping (patience = 10) on a batch size of 32. The dataset is split into training (70\%), validation (15\%), and test (15\%). Detailed evaluation metrics are provided in Section~\ref{sec:evaluation}.

\begin{figure}[t]
\centering
\subfloat[Home screen]{
  \fbox{\includegraphics[width=0.44\columnwidth, height=4cm]{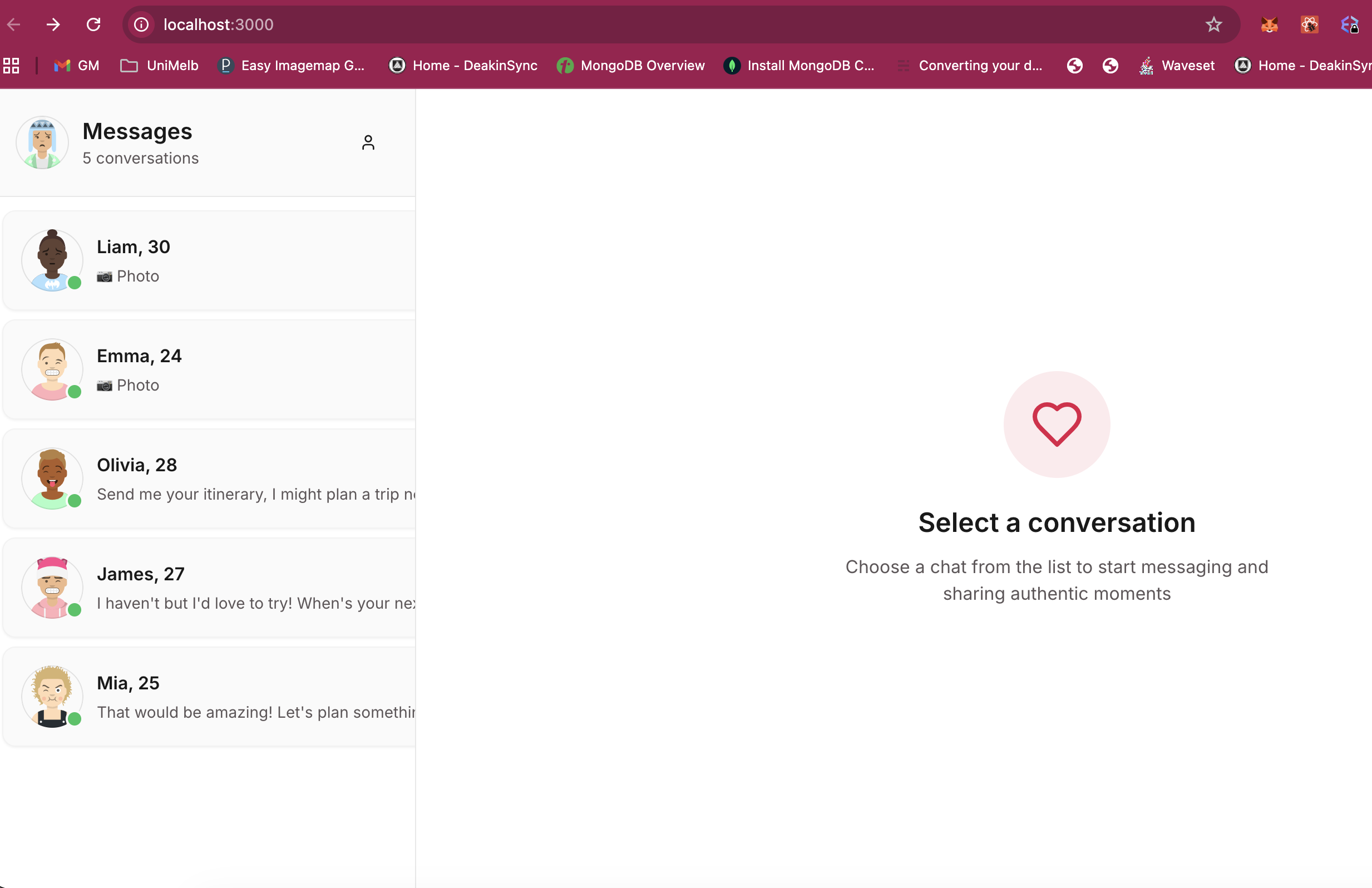}}
  \label{fig:home}}
\hfill
\subfloat[Image being validated]{
  \fbox{\includegraphics[width=0.44\columnwidth, height=4cm]{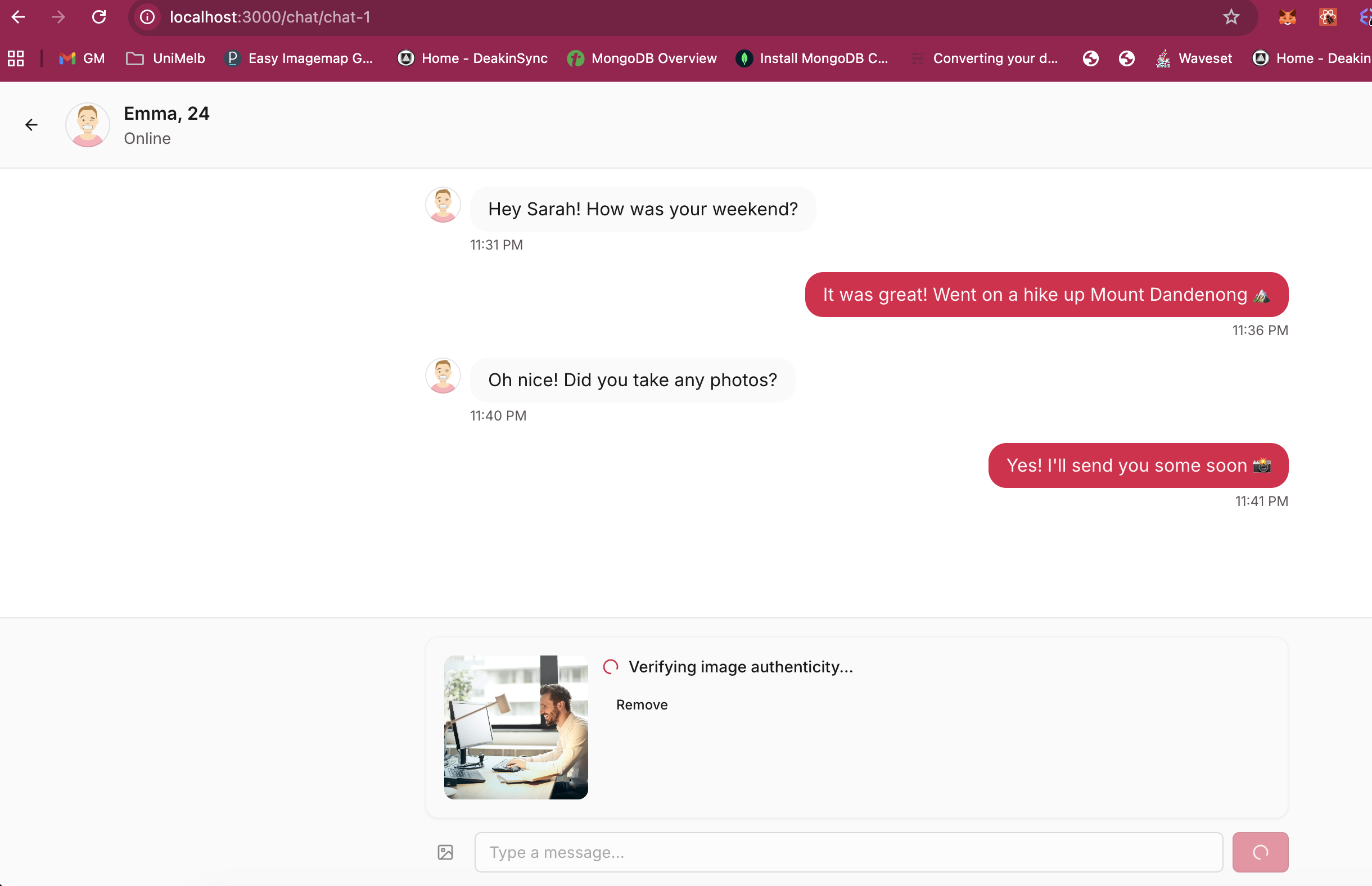}}
  \label{fig:validating}}
\caption{The SecureChatDate application. (a)~Home screen listing 
active conversations with message previews and timestamps. 
(b)~An uploaded image undergoing validation with the send button 
disabled until classification completes.}
\label{fig:app-overview}
\end{figure}

\begin{figure}[t]
\centering
\subfloat[Original image accepted]{
  \fbox{\includegraphics[width=0.45\columnwidth,  height=4cm]{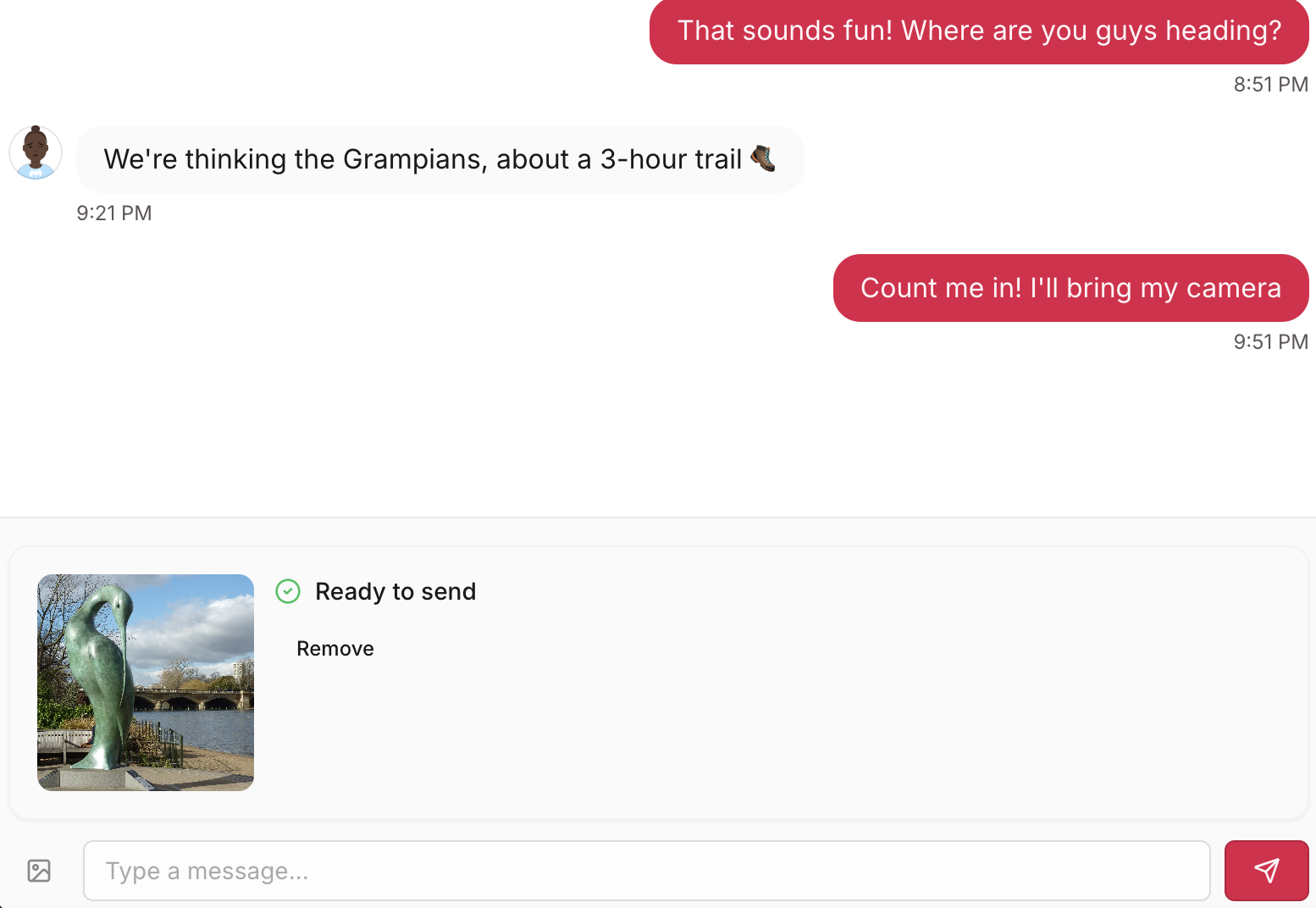}}
  \label{fig:accepted}}
\hfill
\subfloat[Recaptured image blocked]{
  \fbox{\includegraphics[width=0.45\columnwidth,  height=4cm]{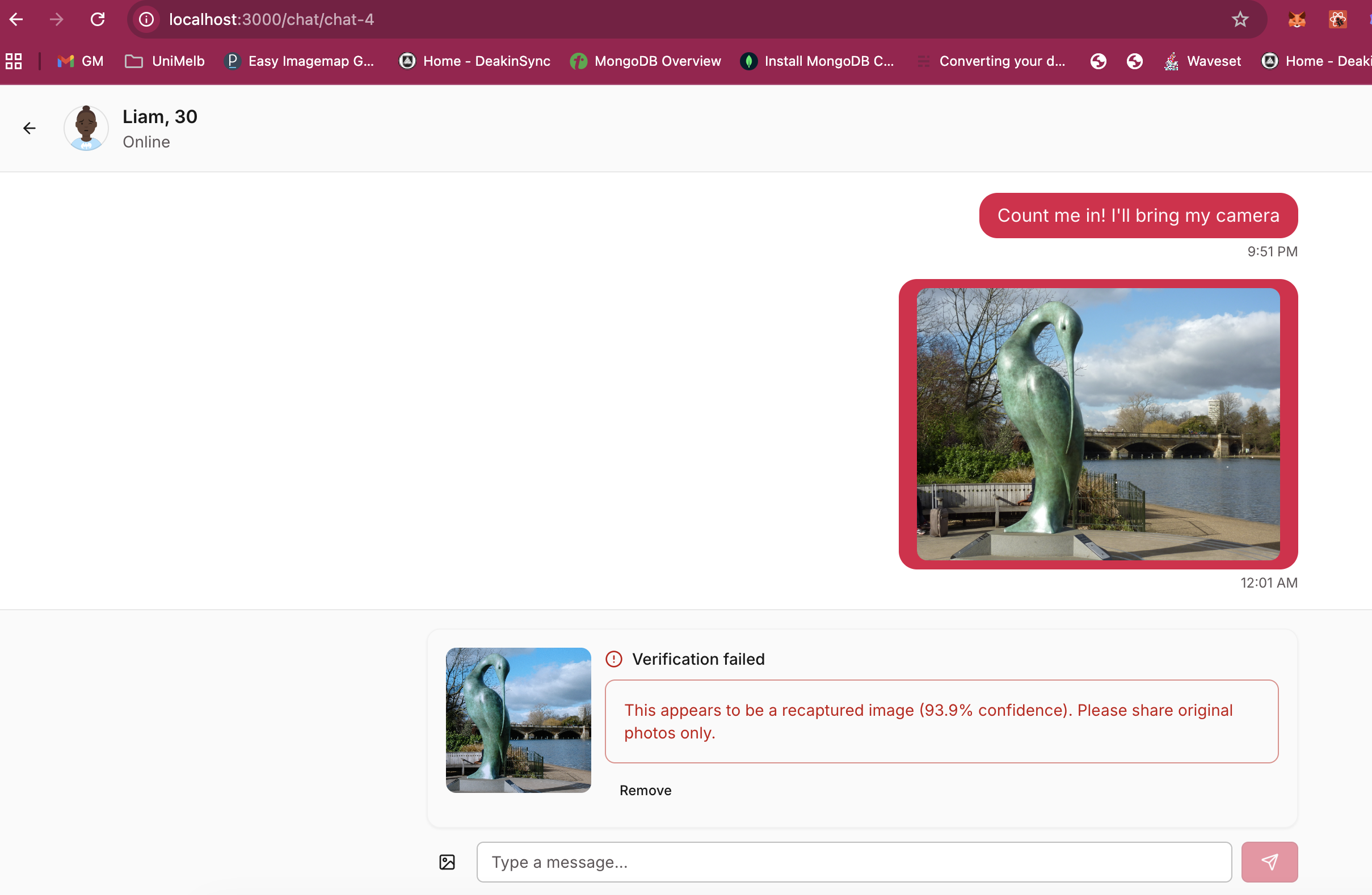}}
  \label{fig:blocked}}
\caption{Client-side enforcement in our proof of concept SecureChatDate application. 
(a)~An original photograph passes validation with high confidence, 
enabling the send button. 
(b)~A recaptured image is detected and blocked, displaying the 
confidence score to the user.}
\label{fig:app-enforcement}
\end{figure}

\section{\textbf{SecureChatDate} Web Application: A Client-Side Enforcement Mechanism}
\label{sec:web-application}

To demonstrate practical applicability, we developed a proof-of-concept dating chat application named as \textit{SecureChatDate} with integrated image authentication that enforces a zero-trust policy---all shared images undergo mandatory validation before transmission. The application employs a three-tier architecture: a React/TypeScript frontend providing real-time chat via WebSocket with image upload and validation feedback; a Node.js/Express backend exposing two endpoints---\texttt{/api/validate-image} (classification) and \texttt{/api/upload} (storage of validated images only)---with immediate deletion of temporary files; and a PyTorch inference engine that loads the EE-CNN weights once at initialization and returns JSON results containing prediction label, confidence score, and class probabilities. Fig.~\ref{fig:app-overview} shows the application interface, including the conversation list and the image validation process. %\textbf{Enforcement Workflow:-} 
The enforcement pipeline is as follows:

\begin{enumerate}
    \item \textbf{User action}: User selects an image file from their device.
    \item \textbf{Client validation trigger}:
    \begin{itemize}
        \item File type check (must be \texttt{image/*}) and size validation (maximum 10MB)
        \item Local preview generated; send button disabled until validation completes.
    \end{itemize}
    \item \textbf{Server validation}:
    \begin{itemize}
        \item Image uploaded to \texttt{/api/validate-image}
        \item Server invokes inference script using following python command: \\ \texttt{python3 inference.py model.pth image.jpg}
        \item CNN processes image and returns JSON result
    \end{itemize}
    \item \textbf{Decision point}:
    \begin{itemize}
        \item \textit{If Original ($p_{\text{orig}} > p_{\text{recap}}$)}: Server returns \texttt{\{isValid: true\}}; client displays green checkmark and enables send button
        \item \textit{If Recaptured ($p_{\text{recap}} > p_{\text{orig}}$)}: Server returns \texttt{\{isValid: false\}} with confidence score; image deleted from server; client displays error with blocking message; send button remains disabled.
        \item \textit{If model unavailable}: System defaults to blocking (fail-closed policy)
    \end{itemize}
    \item \textbf{Image transmission} (only if validated):
    \begin{itemize}
        \item Image uploaded to \texttt{/api/upload}; server stores and generates permanent URL.
        \item WebSocket broadcast notifies the recipient in real-time.
    \end{itemize}
\end{enumerate}

Fig.~\ref{fig:app-enforcement} demonstrates the enforcement mechanism in the proof-of-concept application, showing both acceptance and rejection outcomes.

\subsection{Proposed Integration with Mobile Chat Applications}

Beyond the web-based PoC, we outline how the framework can be integrated with native mobile chat applications to enforce recapture detection. % across all image-sharing pathways. \par
\par 
\textbf{1. Android integration:-} The framework can intercepts image-sharing at the OS level through three mechanisms:
\begin{itemize}
    \item \textbf{Share-intent interception:-} An \texttt{IntentFilter} for \texttt{ACTION\_SEND} with MIME type \texttt{image/*} enables a background validation service to intercept outgoing shares, forwarding the intent only if the image passes validation.
    
    \item \textbf{Accessibility service:-} An \texttt{AccessibilityService} monitors file picker interactions across protected applications, triggering validation before share actions complete (requires explicit user consent).
    
    \item \textbf{Background service and on-device inference:-} The model runs as a persistent \texttt{ForegroundService} using a TensorFlow Lite model ($\sim$9MB, INT8 quantized). Inference is dispatched via NNAPI to available accelerators (GPU, DSP, or NPU), achieving 60--150ms latency. A \texttt{ContentObserver} on \texttt{MediaStore.Images} proactively classifies new gallery images before sharing.
\end{itemize}

\textbf{2. iOS Apple:-} %On iOS, 
The framework operates within Apple's sandboxing model:
\begin{itemize}
    \item \textbf{Share extension:-} Intercepts image-sharing actions system-wide for \texttt{public. image} UTI, loading the Core ML model to classify and either forward or block content within Apple's 120MB extension memory limit.
    
    \item \textbf{PHPickerViewController hook:-} Validation is injected into the picker's completion handler, classifying selected \texttt{PHAsset} objects before the application receives image data.
    
    \item \textbf{On-device inference via core ML:-} The model executes on Apple's Neural Engine (40--100ms on A14 Bionic or later), with state shared across extensions via \texttt{App Groups} containers.
\end{itemize}

\textbf{3. Cross-platform implementation:-} Both platforms share three mechanisms: (1)~over-the-air model updates without full app releases, (2)~local SQLite caching of validation results keyed by SHA-256 content hash, and (3)~fully on-device inference %with no network dependency, 
preserving zero-trust operation in offline scenarios.

\section{IMI for Forensic Traceability}
\label{sec:IMI}

A complementary forensic layer is needed to trace leakage when 
recaptured content evades detection. We introduce the concept of 
an IMI  lightweight, 
imperceptible marker embedded into original images before 
transmission. The IMI must satisfy three properties: 
\textit{invisibility} (imperceptible to the human eye), 
\textit{survivability} (persists through the screen-camera 
recapture process, unlike EXIF metadata), and 
\textit{traceability} (encodes sufficient information to link 
a leaked image to its source). When an original image $I$ passes validation 
(Algorithm~\ref{alg:enforcement}, Stage~4), the IMI module 
encodes a payload $\mathcal{P} = \langle \text{userID}, 
\text{timestamp}, \text{sessionID} \rangle$ by targeting 
mid-frequency DCT coefficients robust to display-camera 
distortion. The embedded image is:

\begin{equation}
I_{\text{marked}} = I + \alpha \cdot \mathcal{W}(\mathcal{P})
\end{equation}

\noindent where $\mathcal{W}(\cdot)$ maps the payload to a 
spatial-domain perturbation and $\alpha$ balances invisibility 
against recapture survivability. If a leaked image is discovered, 
inverse frequency-domain analysis recovers the payload to 
identify the original recipient and establish the leakage path.

This component is explored at a conceptual level, informed by recent works in watermarking: Bai~\textit{et al.}~\cite{bai2022ssden} and Cao~\textit{et al.}~\cite{cao2024universal} demonstrated that deep watermarks survive the screen-camera channel, suggesting that embedding a compact 64--128 bit payload is feasible. Further, we have in our previous work~\cite{yu2014effective} demonstrated the feasibility of  packet marking using mobile International Mobile Equipment Identity (IMEI number, a mobile hardware number) as unique marks for the traceback purpose.  Many open challenges, aim to address in the future, include maintaining embedding latency under 100ms on mobile, surviving  adversarial post-processing, and GDPR compliance for user-identifying information.

\section{Experimental Evaluation}
\label{sec:evaluation}

%We evaluate the proposed Edge-Enhanced CNN on quantitative metrics and qualitative visualizations to assess its effectiveness in distinguishing original photographs from recaptured images.

We employ standard classification metrics: accuracy, precision, recall, F1-score, average confidence, and high-confidence accuracy (predictions with confidence $> 0.8$). We additionally analyze the ROC curve and AUC across classification thresholds. The held-out test set comprises 450 images (225 original, 225 recaptured) captured with varied smartphones (iPhone 12/13, Samsung Galaxy S21/S22, Google Pixel 6) and recaptured from diverse displays (OLED, AMOLED, Retina, LCD) under systematically varied conditions (indoor/outdoor lighting, angles $15^\circ$--$45^\circ$, distances 20--50cm).
%\subsection{Quantitative Results}
Table~\ref{tab:performance} summarizes the performance.
\begin{table}[h]
\centering
\caption{Model's performance metrics on the held-out test set (450 images)}
\label{tab:performance}
\begin{tabular}{lc }
\hline
\textbf{Metric} & \textbf{Value}  \\
\hline
Overall Accuracy & 98.89\% \\
Precision (Original) & 97.83\% \\
Recall (Original) & 99.11\% \\
F1-Score & 98.46\% \\
Average Confidence & 94.57\% \\
High-Confidence Accuracy & 99.46\% \\
\hline
False Positive Rate & 1.78\% \\
False Negative Rate & 0.89\% \\
\hline
\end{tabular}
\end{table}

\textbf{Analysis of results:-} The EE-CNN achieves \textbf{98.89\% accuracy}, correctly classifying 445 out of 450 test images. High precision (97.83\%) ensures legitimate users are rarely blocked, while recall of 99.11\% provides strong protection against recaptured content. When restricted to high-confidence predictions (confidence $> 0.8$), accuracy rises to \textbf{99.46\%}, with only 2--3 misclassifications, confirming that model uncertainty correlates well with prediction difficulty. The FPR of 1.78\% (approximately 4 of 225 original images) arises primarily from images with extreme motion blur, photographs taken through glass or reflective surfaces, and images with heavy prior compression artifacts. The FNR of 0.89\% (approximately 2 of 225 recaptured images) occurs when recapture is performed with high-end OLED displays at near-perpendicular angles under ambient lighting that closely matches display brightness, minimizing moir\'e effects. These edge cases approach the theoretical limits of recapture detection.\par 

\begin{figure}[h]
\centering
\includegraphics[width=0.5\columnwidth]{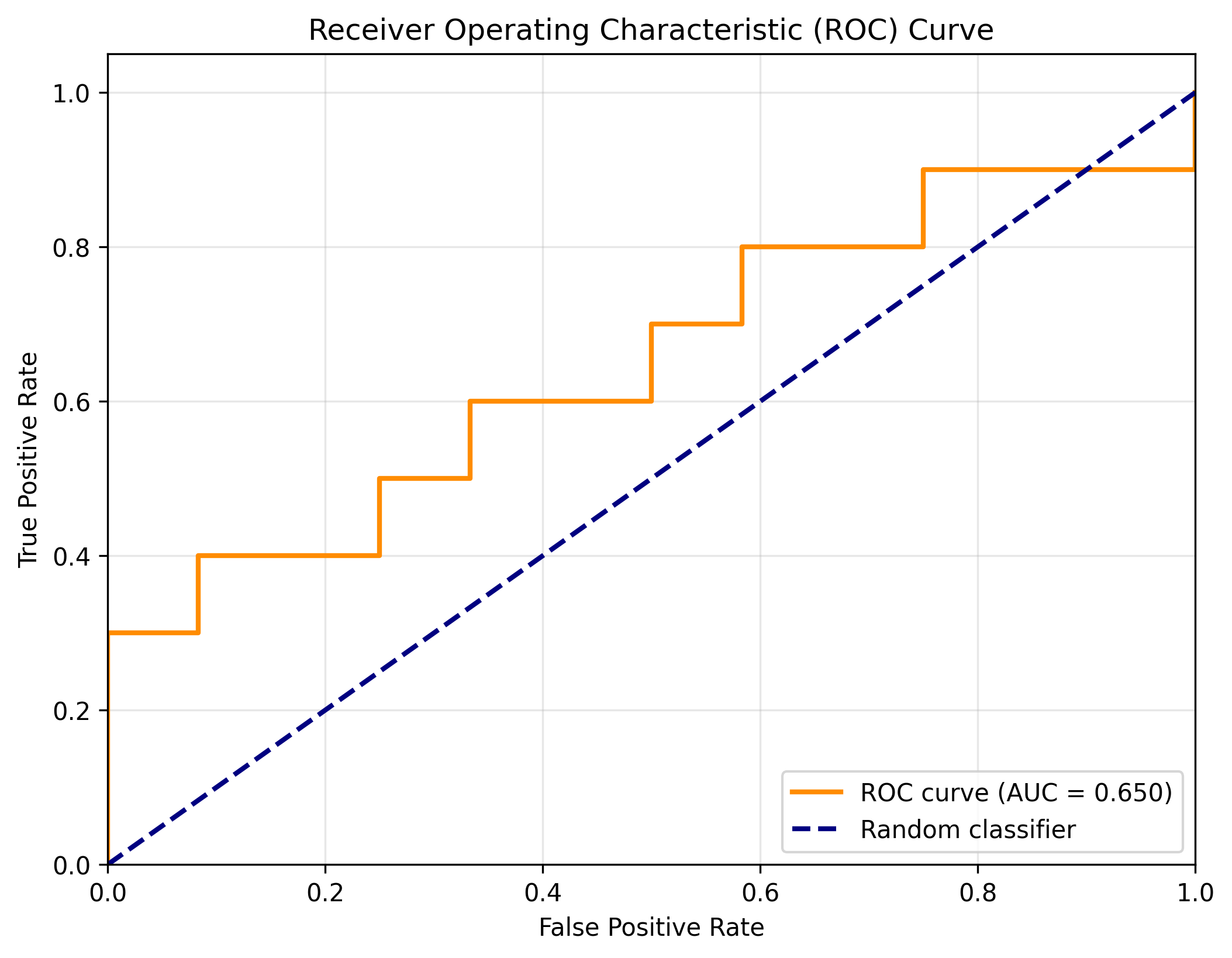}
\caption{Training and validation loss (left) and accuracy (right) over 50 epochs. The model converges rapidly within the first 15 epochs and maintains stable performance thereafter, with minimal overfitting.}
\label{fig:training-curves}
\end{figure}

\textbf{Training dynamics:-} Fig.~\ref{fig:training-curves} illustrates the training and validation loss and accuracy over 50 epochs.
Both training and validation accuracy surge from approximately 60\% to 90\% within the first 5 epochs---indicating strong initial representations from the edge enhancement layer---and stabilize around epoch 15--20, with validation accuracy slightly exceeding training accuracy (94--95\% vs.\ 92--93\%), confirming that dropout, batch normalization, and data augmentation effectively prevent overfitting. Beyond epoch 20, performance remains stable ($\pm 1\%$ fluctuation), validating early stopping at epoch 30--40 as sufficient.\par 

\begin{figure*}[t]
\centering
\includegraphics[width=\textwidth]{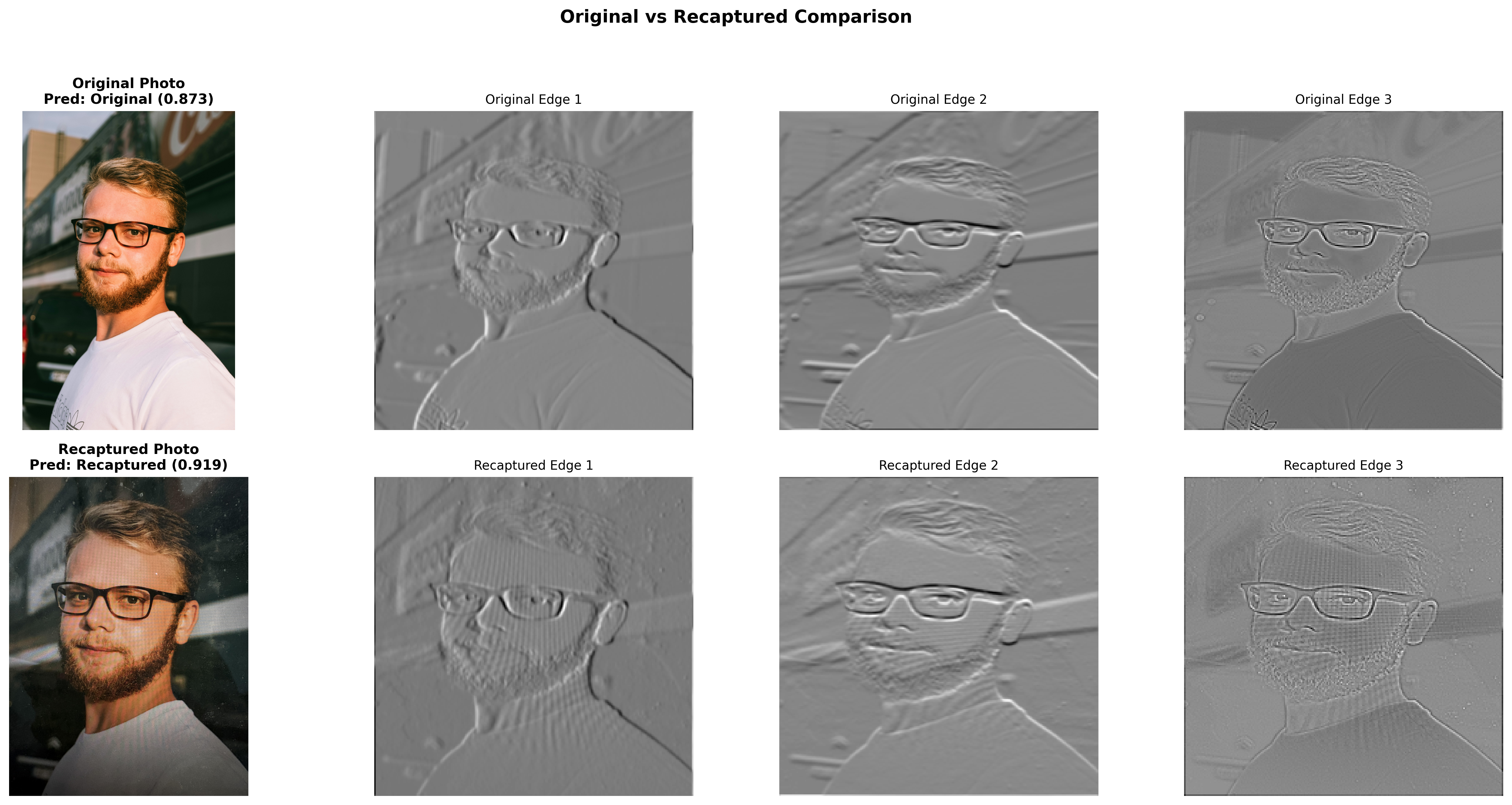}
\caption{Edge detection filter responses comparing original (top row) and recaptured (bottom row) images. Sobel-X (left), Sobel-Y (middle), and Laplacian (right) filters. Recaptured images exhibit moir\'e patterns, grid-like interference from the pixel matrix, and inconsistent edge sharpness.}
\label{fig:edge-comparison}
\end{figure*}

\textbf{Qualitative analysis (edge detection effectiveness):-} To understand how the model distinguishes original from recaptured images, we visualize the activations of the edge enhancement layer and deeper convolutional blocks. Fig.~\ref{fig:edge-comparison} shows the edge filter responses for an original image and its recaptured counterpart. Original images produce clean, sharp edge transitions corresponding to actual scene boundaries with consistent intensity gradients. Recaptured images, by contrast, exhibit several distinctive artifacts in the edge domain: moir\'e patterns (wavy, repetitive distortions from sensor-display pixel grid interference, prominent in Sobel responses), grid artifacts (faint structures from the display pixel matrix, visible in Laplacian responses), softened edge transitions from the two-step capture-display-recapture pipeline, and amplified high-frequency noise from display refresh and ambient lighting interference.\par

\begin{figure*}[t]
\centering
\includegraphics[width=\textwidth]{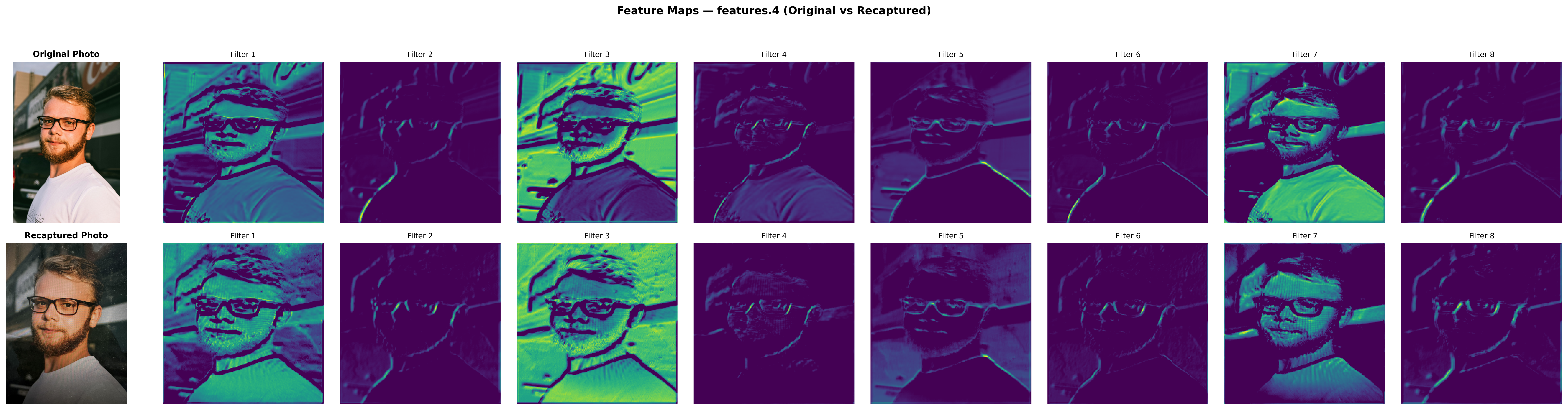}

{\small \textbf{(a)} Block 1 (features.4) — Early layer features responding to edge orientations, texture gradients, and local contrast.}

\vspace{1em}

\includegraphics[width=\textwidth]{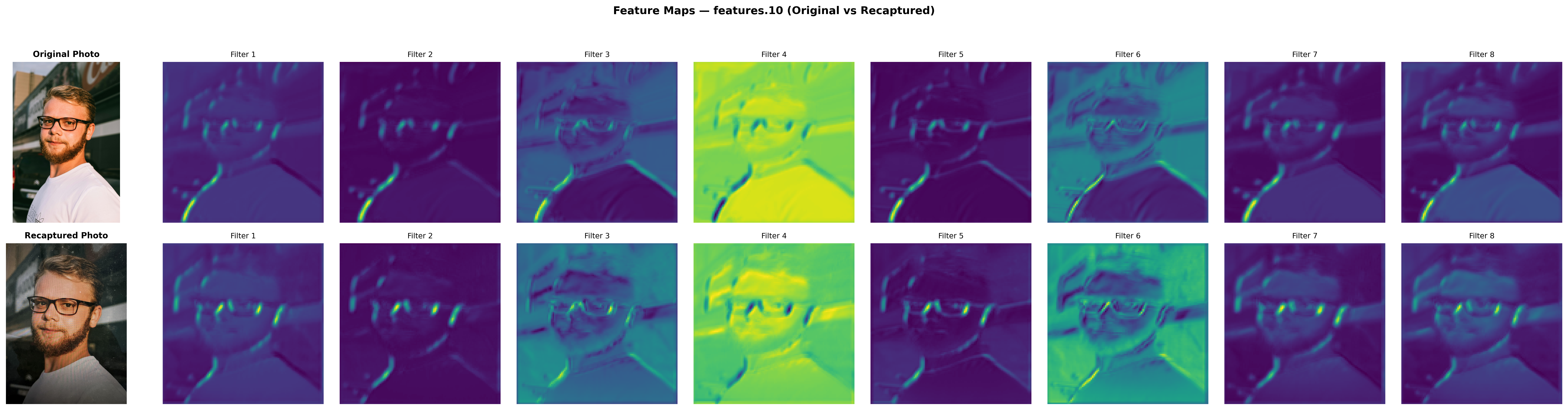}

{\small \textbf{(b)} Block 2 (features.10) — Mid-layer pattern detectors capturing moir\'e structures, grid formations, and compound edge-texture signatures.}

\caption{Feature map visualizations from convolutional blocks.}
\label{fig:feature-maps}
\end{figure*}

Fig.~\ref{fig:feature-maps} displays feature map activations from deeper convolutional blocks, revealing how these edge-level cues are progressively abstracted into higher-order discriminative features. Early-layer feature maps (Block~1) respond to basic edge orientations and local contrast variations, with filters 4, 5, 9, and 10 showing strong selective activation on recaptured images, suggesting specialization for high-frequency artifact detection. Mid-layer maps (Block~2) capture more abstract patterns---repetitive moir\'e structures, grid-like formations, and compound edge-texture signatures---with diverse, complementary activations confirming that the network avoids redundant representations. \par 

\textbf{Comparison with baseline:-} We compare against a ResNet-18 baseline (ImageNet pre-trained, fine-tuned on our dataset) without edge enhancement, see Table~\ref{tab:comparison}.
\begin{table}[h]
\centering
\caption{Comparison with ResNet-18 baseline}
\label{tab:comparison}
\begin{tabular}{lcc}
\hline
\textbf{Metric} & \textbf{ResNet-18} & \textbf{EE-CNN (Ours)} \\
\hline
Accuracy & 94.22\% & \textbf{98.89\%} \\
Precision & 92.15\% & \textbf{97.83\%} \\
Recall & 96.44\% & \textbf{99.11\%} \\
F1-Score & 94.24\% & \textbf{98.46\%} \\
Parameters & 11.7M & 8.7M \\
Inference Time (CPU) & 680ms & \textbf{450ms} \\
\hline
\end{tabular}
\end{table}
The EE-CNN outperforms ResNet-18 by \textbf{4.67 percentage points} in accuracy while using 25\% fewer parameters and achieving 34\% faster inference, demonstrating that the edge enhancement layer provides a strong inductive bias for recapture detection compared to generic feature extractors. While the model achieves strong performance, several limitations warrant consideration: (1)~near-perfect recaptures using professional displays and controlled lighting can occasionally evade detection (FNR = 0.89\%); (2)~original images with extensive post-processing may trigger false positives (FPR = 1.78\%); and (3)~the model is trained primarily on smartphone-to-screen-to-smartphone recaptures, and performance on other scenarios (DSLR cameras, printed photos) requires further validation. Future work should explore multi-scale analysis at different spatial frequencies, frequency-domain augmentation via Fourier or wavelet features, adversarial robustness evaluation, and attention-based explainability mechanisms.

\section{Security Analysis and Discussion}
\label{sec:discussion}

%\subsection{Security Architecture}
%The user uploads an image to the server and the EE-CNN classifies it with a confidence score. The enforcement logic accepts the image if predicted original with confidence $> 0.80$, rejects it if predicted recaptured with confidence $< 0.80$, or routes it to a manual review queue otherwise. %The rejected users receive guidance on submitting original photographs. 
%The model executes server-side for four reasons: tamper resistance (client-side models can be reverse-engineered), centralized updates (improvements deploy without app updates), audit trail (logging enables forensic analysis), and resource control (consistent GPU acceleration regardless of user device).\par 

\textbf{1. Resistance to bypass attempts:-}\par 
\textbf{I. Image post-processing attacks:-} Adversaries may apply Gaussian blur, histogram equalization, or adaptive sharpening to suppress recapture artifacts. The EE-CNN's edge enhancement layer captures structural distortions---doubled edges, frequency-domain aliasing, and illumination gradients---that persist after moderate filtering. Applying Gaussian blur ($\sigma = 1.5$) and contrast enhancement ($\pm 15\%$) to recaptured test images reduced detection accuracy by only 2--3\%. \par \textbf{II. Format and compression manipulation:-} Re-encoding through multiple JPEG compression cycles or format conversion may alter forensic traces. The training pipeline includes JPEG compression at quality levels 70--95 as augmentation, and the EE-CNN operates on spatial-domain edge responses rather than compression-specific artifacts, making its discriminative power largely format-independent. \par 
\textbf{III. API-Level and protocol attacks:-} Adversaries might bypass detection by directly injecting image data into the messaging pipeline. The backend enforces mandatory validation---no image reaches the chat storage or WebSocket layer without a passing classification. API authentication tokens and rate limiting (10 requests per minute per user) prevent automated threshold probing.\par 

\textbf{2. False positive analysis and mitigation:-} The system achieves an FPR of 1.78\%, corresponding to approximately 18 false rejections per 1,000 legitimate uploads. False positives concentrate in three scenarios: images under fluorescent/LED lighting (flicker resembling screen backlighting), images with reflective surfaces (specular highlights mimicking screen recapture), and close-range images with shallow depth of field (bokeh resembling screen-camera focus degradation). The framework addresses false positives through: \textbf{Confidence thresholding}---images with confidence $\leq 0.80$ are routed to manual review rather than outright rejection; \textbf{User feedback}---rejected users receive explanatory notifications with confidence scores and retry guidance; \textbf{EXIF Cross-validation}---borderline cases with consistent, unmodified EXIF metadata are escalated for review rather than auto-rejected; and \textbf{Continuous retraining}---false positive cases from the review queue are incorporated into subsequent training cycles.\par 

\textbf{3. Performance overhead and  scalability:-} The EE-CNN achieves 85ms inference on GPU (NVIDIA Tesla V100) and 450ms on CPU. The other metrics are shown in % End-to-end API validation latency averages 800--1,200ms over 4G. The model occupies $\sim$35MB serialized, with peak memory of $\sim$200MB per concurrent request, allowing 8--12 concurrent inferences per GPU instance. 
Table~\ref{tab:performance1}. % shows the performance characteristics.\par 
 Further, the architecture supports horizontal scaling via containerized deployment (Docker/Kubernetes) with auto-scaling policies. Model quantization (FP32 $\rightarrow$ INT8) can reduce inference time by 2--3$\times$ with $<$0.5\% accuracy drop, and on-device deployment via TensorFlow Lite or Core ML eliminates server round-trips for latency-critical applications.\par 

\begin{table}[H]
\centering
\caption{Performance overhead of the detection pipeline}
\label{tab:performance1}
\begin{tabular}{|l|c|}
\hline
\textbf{Metric} & \textbf{Value} \\
\hline
Model size (serialized) & $\sim$35 MB \\
\hline
GPU inference latency & 85 ms \\
\hline
CPU inference latency & 450 ms \\
\hline
End-to-end API latency (4G) & 800--1,200 ms \\
\hline
Peak memory per request & $\sim$200 MB \\
\hline
Concurrent requests per GPU & 8--12 \\
\hline
Preprocessing time & $\sim$15 ms \\
\hline
\end{tabular}
\end{table}

\textbf{4. Privacy and ethical considerations:-} The system processes images solely for authentication without extracting biometric identifiers---no facial recognition, identity matching, or demographic inference is performed. Images are analyzed in-memory and discarded after classification. Device bias (older smartphones triggering false positives) and environmental bias (challenging lighting causing false rejections) are mitigated through diverse data processing.% (training data and regular bias audits).

\section{Conclusion}
\label{sec:conclusion}

We presented a novel \textit{screen recaptured analog hole attack} (S-RAHA) that circumvents conventional screenshot-prevention mechanisms. We proposed a secure-by-design framework for detecting and preventing the forwarding of re-captured images. The framework integrates an EE-CNN model that exploits recapture-specific artifacts---moir\'e patterns, edge degradation, and illumination non-uniformity---achieving 98.89\% detection accuracy %with a false positive rate of 1.78\% 
on a diverse test set spanning multiple device
and environmental configurations. The proposal outperforms a ResNet-18 baseline by 4.67\% points in accuracy while requiring 25\% fewer parameters and 34\% faster inference, demonstrating the effectiveness of the proposal. %domain-specific edge enhancement for recapture detection. 
The framework incorporates a client-side enforcement mechanism that blocks sharing of suspected recaptured images before transmission, operating under a zero-trust model with no
dependency on network connectivity or server-side validation. A
proof-of-concept web application we built %with React, Express.js, and
%PyTorch 
demonstrates the practical feasibility of the approach. 
%, while the architecture generalises to native mobile platforms through
%TensorFlow Lite and Core ML. 
We further introduced the concept of an invisible metadata identifier for forensic traceability of leakage paths, explored at a feasibility level as a complementary
forensic layer. While limitations remain---particularly around high-fidelity recaptures from modern OLED displays and the need for larger, more diverse training datasets---the framework establishes a practical foundation for integrating recapture detection into privacy-focused web and mobile applications. Future work focus on expanding the training corpus, incorporating frequency-domain features, and evaluating adversarial robustness. %, and developing a full native mobile implementation with on-device inference.

%\section*{Acknowledgments}
%This should be a simple paragraph before the References to thank those individuals and institutions who have supported your work on this article.

% {\appendix[Proof of the Zonklar Equations]
% Use 

\bibliographystyle{IEEEtran}
\bibliography{references}

@article{ke2013multiple,
  title={Image recapture detection using multiple features},
  author={Ke, Qingshan and Li, Xiaoshuai and Shi, Yong and Yan, Qiong},
  journal={International Journal of Multimedia and Ubiquitous Engineering},
  volume={8},
  number={4},
  pages={101--114},
  year={2013}
}

@article{thongkamwitoon2015edge,
  author={Thongkamwitoon, Thirapiroon and Muammar, Hani and Dragotti, Pier-Luigi},
  journal={IEEE Transactions on Information Forensics and Security},
  title={An Image Recapture Detection Algorithm Based on Learning Dictionaries
         of Edge Profiles},
  year={2015},
  volume={10},
  number={5},
  pages={953--968}
}

@InProceedings{dirik2011smartphone,
  author="Gao, Xinting and Qiu, Bo and Shen, JingJing
          and Ng, Tian-Tsong and Shi, Yun Qing",
  editor="Kim, Hyoung-Joong and Shi, Yun Qing and Barni, Mauro",
  title="A Smart Phone Image Database for Single Image Recapture Detection",
  booktitle="Digital Watermarking",
  year="2011",
  publisher="Springer Berlin Heidelberg",
  address="Berlin, Heidelberg",
  pages="90--104"
}

@InProceedings{luo2021scale,
  author="Luo, Jinian and Guo, Jie and Qiu, Weidong
          and Huang, Zheng and Hui, Hong",
  editor="Mantoro, Teddy and Lee, Minho and Ayu, Media Anugerah
          and Wong, Kok Wai and Hidayanto, Achmad Nizar",
  title="Scale Invariant Domain Generalization Image Recapture Detection",
  booktitle="Neural Information Processing",
  year="2021",
  publisher="Springer International Publishing",
  address="Cham",
  pages="75--86"
}

@article{hussain2025fewshot,
  title={Few-shot based learning recaptured image detection with multi-scale
         feature fusion and attention},
  journal={Pattern Recognition},
  volume={161},
  pages={111248},
  year={2025},
  issn={0031-3203},
  author={Israr Hussain and Shunquan Tan and Jiwu Huang}
}

@Article{bai2022ssden,
  author  = {Bai, Rui and Li, Li and Zhang, Shanqing
             and Lu, Jianfeng and Chang, Chin-Chen},
  title   = {{SSDeN}: Framework for Screen-Shooting Resilient Watermarking
             via Deep Networks in the Frequency Domain},
  journal = {Applied Sciences},
  volume  = {12},
  year    = {2022},
  number  = {19},
  article-number = {9780}
}

@article{cao2024universal,
  title={Universal screen-shooting robust image watermarking with
         channel-attention in {DCT} domain},
  journal={Expert Systems with Applications},
  volume={238},
  pages={122062},
  year={2024},
  issn={0957-4174},
  doi={10.1016/j.eswa.2023.122062},
  author={Fang Cao and Daidou Guo and Tianjun Wang
          and Heng Yao and Jian Li and Chuan Qin}
}

@article{liu2025screenshooting,
  author    = {Lianshan Liu and Peng Xu and Qianwen Xue},
  title     = {Screen shooting resistant watermarking based on cross attention},
  journal   = {Scientific Reports},
  volume    = {15},
  number    = {1},
  pages     = {17016},
  year      = {2025},
  publisher = {Nature Publishing Group},
  doi       = {10.1038/s41598-025-00912-8},
  issn      = {2045-2322}
}

@article{aharon2006ksvd,
  author  = {Aharon, Michal and Elad, Michael and Bruckstein, Alfred},
  title   = {{K-SVD}: An Algorithm for Designing Overcomplete Dictionaries
             for Sparse Representation},
  journal = {IEEE Transactions on Signal Processing},
  volume  = {54},
  number  = {11},
  pages   = {4311--4322},
  year    = {2006},
  doi     = {10.1109/TSP.2006.881199}
}

@inproceedings{chen2024cma,
  title={CMA: a chromaticity map adapter for robust detection of screen-recapture document images},
  author={Chen, Changsheng and Lin, Liangwei and Chen, Yongqi and Li, Bin and Zeng, Jishen and Huang, Jiwu},
  booktitle={Proceedings of the IEEE/CVF Conference on Computer Vision and Pattern Recognition},
  pages={15577--15586},
  year={2024}
}

@article{chen2025moire,
  title={Moire spectral augmentation and masked frequency modeling for document presentation attack detection},
  author={Chen, Changsheng and Li, Youjie and Li, Bokang and Yu, Weifan and Chen, Baoying and Li, Bin and Huang, Jiwu},
  journal={IEEE Transactions on Dependable and Secure Computing},
  year={2025},
  publisher={IEEE}
}

@inproceedings{park2025chimera,
  title={Chimera: Creating Digitally Signed Fake Photos by Fooling Image Recapture and Deepfake Detectors},
  author={Park, Seongbin and Vilesov, Alexander and Zhang, Jinghuai and Khalili, Hossein and Tian, Yuan and Kadambi, Achuta and Sehatbakhsh, Nader},
  booktitle={34th USENIX Security Symposium (USENIX Security 25)},
  pages={4305--4324},
  year={2025}
}

@inproceedings{ji2025towards,
  title={Towards explainable fake image detection with multi-modal large language models},
  author={Ji, Yikun and Hong, Yan and Zhan, Jiahui and Chen, Haoxing and Lan, Jun and Zhu, Huijia and Wang, Weiqiang and Zhang, Liqing and Zhang, Jianfu},
  booktitle={Proceedings of the 33rd ACM International Conference on Multimedia},
  pages={4398--4407},
  year={2025}
}

@inproceedings {279994,
author = {Changjiang Li and Li Wang and Shouling Ji and Xuhong Zhang and Zhaohan Xi and Shanqing Guo and Ting Wang},
title = {Seeing is Living? Rethinking the Security of Facial Liveness Verification in the Deepfake Era},
booktitle = {31st USENIX Security Symposium (USENIX Security 22)},
year = {2022},
isbn = {978-1-939133-31-1},
address = {Boston, MA},
pages = {2673--2690},
url = {https://www.usenix.org/conference/usenixsecurity22/presentation/li-changjiang},
publisher = {USENIX Association},
month = aug
}

@article{li2025comprehensive,
  title={A comprehensive survey of specularity detection: state-of-the-art techniques and breakthroughs},
  author={Li, Fengze and Ma, Jieming and Liang, Hai-Ning and Tian, Zhongbei and Wu, Zhijing and Wen, Tianxi and Liu, Dawei},
  journal={Artificial Intelligence Review},
  volume={58},
  number={7},
  pages={218},
  year={2025},
  publisher={Springer}
}

@article{wang2023coarse,
  title={Coarse-to-fine disentangling demoir{\'e}ing framework for recaptured screen images},
  author={Wang, Ce and He, Bin and Wu, Shengsen and Wan, Renjie and Shi, Boxin and Duan, Ling-Yu},
  journal={IEEE Transactions on Pattern Analysis and Machine Intelligence},
  volume={45},
  number={8},
  pages={9439--9453},
  year={2023},
  publisher={IEEE}
}

@inproceedings{chen2025unmask,
  title={Unmask Tampering: Efficient Document Tampering Localization under Recapturing Attacks with Real Distortion Knowledge},
  author={Chen, Changsheng and Chen, Wenyu and Lin, Yinyin and Li, Bin and Huang, Jiwu},
  booktitle={Proceedings of the 2025 ACM SIGSAC Conference on Computer and Communications Security},
  pages={1694--1708},
  year={2025}
}

@article{li2023recaptured,
  title={Recaptured screen image identification based on vision transformer},
  author={Li, Guihao and Yao, Heng and Le, Yanfen and Qin, Chuan},
  journal={Journal of Visual Communication and Image Representation},
  volume={90},
  pages={103692},
  year={2023},
  publisher={Elsevier}
}

@article{yu2014effective,
  title={An effective and feasible traceback scheme in mobile internet environment},
  author={Yu, Shui and Sood, Keshav and Xiang, Yong},
  journal={IEEE Communications Letters},
  volume={18},
  number={11},
  pages={1911--1914},
  year={2014},
  publisher={IEEE}
}
\end{document}